\pdfoutput=1
\documentclass[%
   aps,%
   twocolumn,%
   prb,%
   groupedaddress,%
   superscriptaddress,%
   showpacs,%
   showkeys,%
  letterpaper,%
   amsfonts,%
   floatfix,%
]{revtex4}

\RequirePackage{amsmath,amssymb,bm}
\RequirePackage{graphicx}
\RequirePackage[FIGTOPCAP,nooneline]{subfigure}
\RequirePackage{hyperref}
\hypersetup{%
  breaklinks = {true},
  citecolor = {blue},
  colorlinks = {true},
  linkcolor = {red},
  pdfauthor = {\textcopyright\ Vitor M. Pereira },
  pdfcreator = {\LaTeX\ and \flqq hyperref\frqq},
  pdffitwindow = {true},
  pdfmenubar = {true},
  pdfpagelayout = {SinglePage},
  pdfstartview = {Fit},
  pdftoolbar = {true},
  plainpages = {false},
}
\usepackage[usenames,dvipsnames]{xcolor}

\newcommand{\intra}{\emph{intra}-band\ }
\newcommand{\inter}{\emph{inter}-band\ }

\newcommand{\avg}[1]{\ensuremath{\langle{#1}\rangle}}
\newcommand{\lv}{\left|}
\newcommand{\rv}{\right|}
\newcommand{\ra}{\right\rangle}
\newcommand{\la}{\left\langle}
\newcommand{\Fref}[1]{Fig.~\ref{#1}}
\newcommand{\Eqref}[1]{Eq.~(\ref{#1})}
\renewcommand{\eqref}[1]{\ref{#1}}

\begin{document}


\title{Enhanced Optical Dichroism of Graphene Nanoribbons}

\author{F. Hipolito}
\affiliation{NUS Graduate School for Integrated Sciences and
                Engineering,
                Centre for Life Sciences, Singapore 117456}
\affiliation{Graphene Research Centre and Department of Physics,
             National University of Singapore,
             2 Science Drive 3, Singapore 117542}

\author{A. J. Chaves}

\author{R. M. Ribeiro}

\author{M. I. Vasilevskiy}
\affiliation{Department of Physics and Centre of Physics,
             University of Minho, Campus of Gualtar,
             4710-057, Braga, Portugal}

\author{Vitor M. Pereira}
\email{vpereira@nus.edu.sg}
\affiliation{Graphene Research Centre and
            Department of Physics, National University of Singapore,
            2 Science Drive 3, Singapore 117542}

\author{N. M. R. Peres}
\email{peres@fisica.uminho.pt}
\affiliation{Department of Physics and Centre of Physics,
             University of Minho, Campus of Gualtar,
             4710-057, Braga, Portugal}
\affiliation{Academy of Sciences of Lisbon,
             R. Academia das Ci\^{e}ncias
             19, 1249-122, Lisboa, Portugal}



\begin{abstract}
The optical conductivity of graphene nanoribbons is analytical and
exactly derived. It is shown that the absence of translational
invariance along the transverse direction allows considerable
intra-band absorption in a narrow frequency window that varies with
the ribbon width, and lies in the THz range domain for ribbons
10--100\,nm wide. In this spectral region the absorption anisotropy
can be as high as two orders of magnitude, which renders the medium
strongly dichroic, and allows for a very high degree of polarization
(up to $\sim85\%$) with just a single layer of graphene. Using a
cavity for impedance enhancement, or a stack of few layer nanoribbons,
these values can reach almost $100\%$. This opens a potential
prospect of employing graphene ribbon structures as efficient
polarizers in the far IR and THz frequencies.
\end{abstract}

\maketitle


%
\section*{Introduction}
Dichroism refers to the ability of some materials to absorb light
differently, depending on the polarization state of the incoming wave,
and leads to effects such as the rotation of the plane of polarization
of light transmitted through them \cite{BW}. This characteristic is
the basis of several elementary optical elements like polarizers, wave
retarders, etc., which are essential building blocks in optics,
photo-electronics and telecommunications. Dichroism, as an intrinsic
property of certain materials and substances, is also widely relevant
for substance characterization in fields ranging from spectroscopy, to
chemistry, to life sciences.

A grid of parallelly aligned metallic wires is a well known textbook
example of a dichroic system, where unpolarized radiation becomes
polarized perpendicularly to the wires, if the wavelength is much
larger than the wire separation \cite{Fizeau:1861}. This example shows
how geometrical anisotropy can be engineered to induce dichroism in
otherwise isotropic media.

Here we unveil the \emph{intrinsic} dichroic properties of graphene
nanoribbons (GNR), and assess how effectively grids of GNRs can be
used as polarizing elements. To our knowledge, the intrinsic
anisotropic absorption characteristics of GNR have not been explored
as we discuss here.

\begin{figure}[t]
  \centering
  \includegraphics*[width=0.48\textwidth]{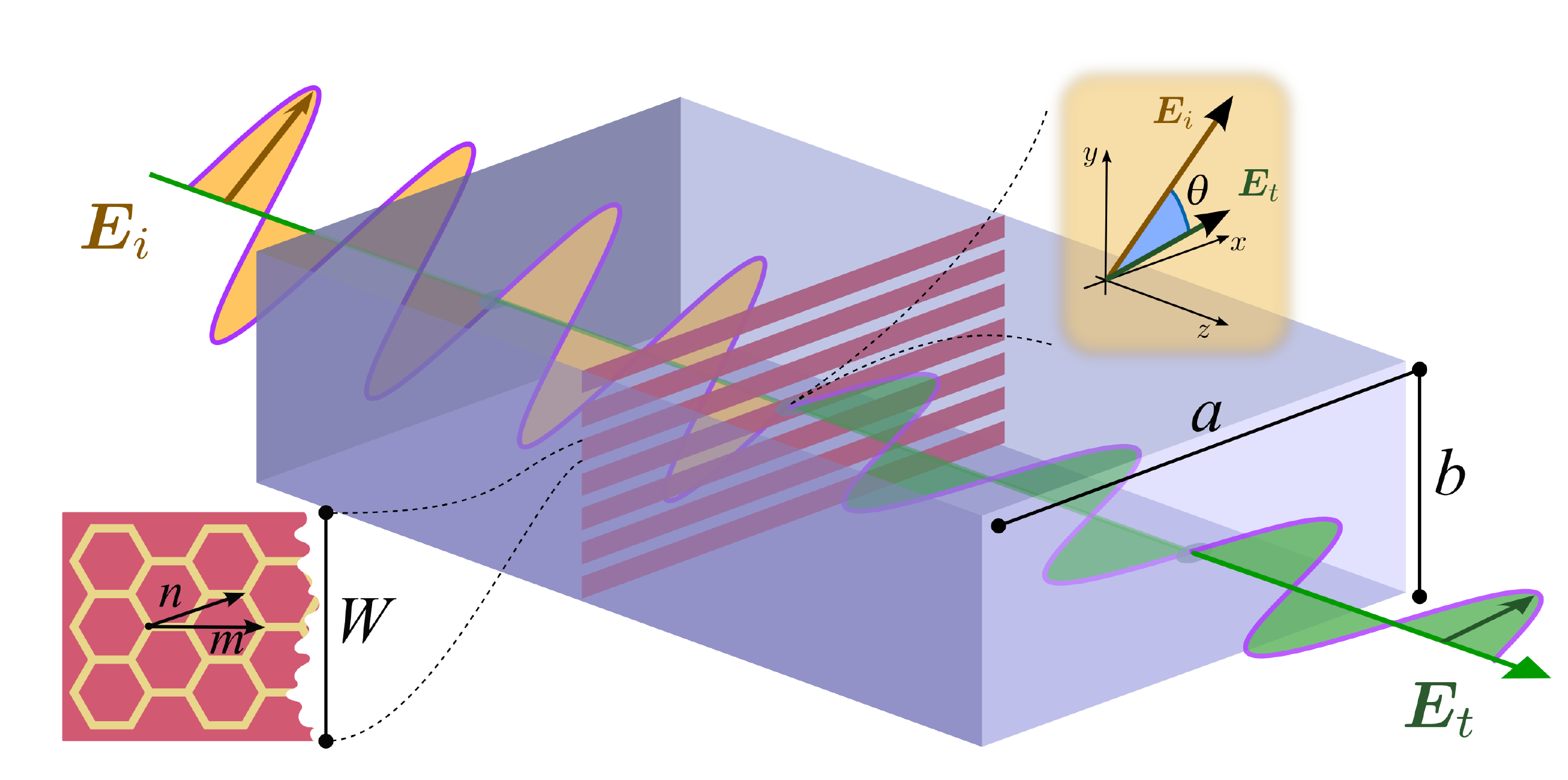}%
  \caption{
    Illustration of the geometry under consideration, and potential
    device application, consisting of a grid of parallel GNRs
    perpendicular to the incoming wave. The grid can be in vacuum, at
    the interface between two different dielectric media (1,2), or
    even inside a metallic waveguide with sectional area $a\times b$.
    A plane-polarized incoming wave has its polarization rotated by an
    angle $\theta$ upon crossing the nanoribbon grating or,
    alternatively, an unpolarized wave emerges linearly polarized.
  }
  \label{fig:Illustration}
\end{figure}

The motivation to explore GNRs in this context comes from a
convergence of several critical properties.
\emph{First}, the optical absorption spectrum of pristine graphene is
roughly constant over an enormous band of
frequencies\cite{Science_Nair,RMP10}, from the THz to the near UV.
This opens the unprecedented prospect of exploring its optical
response to develop optical elements that can operate predictably and
consistently in such broad frequency bands. Broadband polarizers, for
example, are a much needed element in photonic circuits for
telecommunications, and graphene can play here an important role
\cite{KianPing:2011}.
\emph{Second}, the optical absorption of graphene is easily switched
on and off by varying the electronic density, which can be easily
achieved by electrostatic gating \cite{Basov:2008}.
\emph{Third}, due to the record breaking stiffness of the crystal
lattice, one can suspend a graphene sheet and cut a grating of the
thinnest nanowires (currently of the order of 10\,nm
\cite{Lemme:2009}), which opens new avenues in ultra-narrow gratings,
and upon which we base the system depicted in \Fref{fig:Illustration}.
\emph{Fourth}, since graphene is metallic and possesses no bulk (it
is a pure surface), the rich
phenomenology associated with surface plasmons-polaritons (SPP) is
certainly unavoidable, further broadening the horizon of possibilities
for optical applications \cite{Ju2011}.
\emph{Finally}, the atomic thickness of graphene results in a
transparency of 97.7\%. Hence, even if one is able to induce strong
absorption along one direction, the overall transmissivity will still
be large, which is important to maintain losses under control.

%
\section*{Dichroism mechanism}
The natural first step towards such possibilities consists in
analyzing the intrinsic optical response of GNRs, to which we dedicate
the remainder of this paper. We are interested in how the finite
transverse dimension affects the optical absorption spectrum at low
frequencies (IR and below), which is rather featureless in bulk
graphene\cite{Science_Nair} (except for the $\omega=0$ Drude
peak), but turns out to be much richer in nanoribbons. The situation
we envisage is depicted in
\Fref{fig:Illustration}, and consists in passing an electromagnetic
wave across a grid of GNRs. For definiteness and technical simplicity
we restrict our analysis to armchair (AC) nanoribbons, although our
results do not depend on the specific chirality, as will be clear
later.
An important aspect to consider in GNRs has to do with how
large edge disorder is expected to be, and to what extent it might
mask the phenomena under discussion. To address this, while at the
same time keeping as much analytical control over the results as
possible, our calculations involve two steps. First, the
frequency-dependent conductivity tensor $\sigma_{\alpha\beta}(\omega),
\,(\alpha,\beta = x,y)$ of an AC
GNR is derived exactly for free electrons governed by a
nearest-neighbor tight-binding Hamiltonian (see below). We then
perform ensemble averages of such $\sigma_{\alpha\beta}(\omega)$,
where the ribbon width is the fluctuating parameter, and thus extract
the overall response of the system accounting for ``disorder''.
This procedure hinges on the assumption that the leading impact of
disorder in the optical response is captured by the broadening of the
quasi 1D electronic bands, which is also achieved with an ensemble
average of ribbons with fluctuating width. As discussed
below, other generic disorder mechanisms (such as carrier density
inhomogeneity or strain) are supposed to produce only small (of
the order of a few percent) relative fluctuations of the observable
properties of the ribbons. Moreover, such ensemble averaging over a
distribution of ribbon's widths is also close to the experimental
situation, insofar as even state-of-the-art fabrication cannot control
ribbon widths with atomic precision \cite{Xu:2011}. Thus, an array of
ribbons cut out of a graphene sheet will always display a distribution
of widths around a predefined target value $\avg{W}=W_0$. Technically,
the conductivity of such an array of GNRs is given by
$\avg{\sigma_{\alpha\beta}(\omega)}=\sum_{W}f(W)\sigma_{\alpha\beta}
^W(\omega)$, where $f(W)$ is the normal distribution for the ribbon
width $W$, and $\sigma_{\alpha\alpha}^W(\omega)$ is the conductivity
of a single ribbon of width $W$.

Overall parametrizations are as follows.
The natural energy scale is the hopping amplitude in bulk graphene:
$t\simeq 2.7$\,eV \cite{RevModPhys_RevModPhys.81.109}. The chemical
potential, $\mu$, determines the free carrier response and also sets
the spectral limit for \inter transitions (at $T=0$K). Non-zero free
carrier densities are the norm, and their amount depends on the
fabrication and sample treatment procedure. They can range from
$n_e\sim 10^{10}\,\text{cm}^{-2}$ to a few $10^{12}\,\text{cm}^{-2}$.
Such densities correspond to $\mu$ varying roughly between $0.01t$ to
$0.1t$, which is the interval we focus on below. Gating allows the
carrier density to be easily tuned via field effect
\cite{Novoselov:2004}.
Ribbons are interchangeably
characterized by their absolute width $W$, or by $N$, which counts the
number of dimer rows along the transverse direction, and
$W=\sqrt{3}(N-1)a/2\simeq 0.12\,N$\,nm, where $a\simeq 1.42$\,\AA\
represents the C--C distance. For the purposes of ensemble averaging,
ribbon widths are uniformly distributed with a standard deviation that
we take as constant: $\langle N^2 - \avg{N}^2
\rangle^{1/2}=10\,(\simeq1.2\,\text{nm})$. This is done to mimic the
experimental limitations associated with the minimum feature size
that can be achieved by lithographic means, and is presumably a
constant number.
All the calculations discussed below
have been done for $T=300$\,K. We use the terms \emph{intra-} or
\emph{inter-}band in reference to transitions occurring among
subbands with the \emph{same} or \emph{opposite} sign of energy,
respectively. The hopping amplitude sets the energy scale, and all
quantities with dimensions of energy will be expressed in terms of
$t$. For $\mu>0.1t$, and $N>100$ (18\,nm), the finite width of the
ribbon does not significantly alter the relation between $\mu$ and
$n_e$ from the one in bulk graphene. Hence, $n_e \simeq 7\times10^{14}
(\mu/t)^2 \,\text{cm}^{-2}$. To be definite, for illustration
purposes we will take $\mu=0.1t$ in most of the plots\cite{EndNote-5}.
Conductivities are normalized to the universal value $\sigma_0=\pi e^2
/ 2h$ of clean 2D graphene at low frequencies, and the incoming
radiation has a wavelength much larger than the ribbon width $W$.

%
\section*{Derivation of the Conductivity Tensor}
The derivation of the conductivity tensor of an armchair graphene
ribbon starts with the consideration of the nearest neighbor
tight-binding Hamiltonian describing the $\pi$ bands of graphene, and
characterized by a hopping amplitude $t\simeq 2.7$\,eV
\cite{RevModPhys_RevModPhys.81.109}. The ribbon eigenstates have the
analytical form \cite{Katsunori,Zozoulenko}
$
\lv \Psi_{\ell,q,\lambda} \ra = \mathcal{N}
\sum_{n,m}e^{-iq(m+n/2)}\sin \left(k_\ell n\right) \times
  \left(\lv A,n,m\ra + \lambda e^{-i\theta_{\ell,q}} \lv B,n,m\ra
  \right)
$,
where $k_\ell=\pi\ell/(N+1)$ is the quantum number associated with
transverse quantization ($\ell=1,2,\ldots,N$),
$\mathcal{N}=1/\sqrt{N+1}$, $\lambda=\pm 1$ defines the valence
($\lambda=$-1) or conduction ($\lambda=$+1) bands, $\lv A,n,m\ra$ is
the Wannier state at sub-lattice $A$ of the unit cell at position $\bm
R=n\,\bm{n}+m\,\bm{m}$ (see \Fref{fig:Illustration}), $N$ is the
number of unit cells along the finite $\bm{n}$ direction, and $q$ is
the dimensionless momentum along $\bm{m}$, whose value is within the
range $-\pi<q\le\pi$. The phase difference between sub-lattice
amplitudes is
\begin{equation}
  \theta_{\ell,q}=\arctan\frac{2\cos k_\ell\sin\left(q/2\right)}
  {1+2\cos k_\ell\cos\left(q/2\right)}
  \label{eq:theta}
  ,
\end{equation}
This is sufficient to determine the optical conductivity from
Kubo's formula \cite{PRL10}:
\begin{multline}
  \sigma_{\alpha\beta} \!=\! \frac{2ie^2}{\omega S} \!
  \sum_{\ell_1,\ell_2,q}\sum_{\lambda_1,\lambda_2}\!
  \frac{f(E_{\ell_1,q,\lambda_1})-f(E_{\ell_2,q,\lambda_2})}
  {\hbar\omega-(E_{k_2,q,\lambda_2}+E_{k_1,q,\lambda_1})+i0^+}
  \\
  \times \la \Psi_{\ell_1,q,\lambda_1}\rv
  v_{\alpha}\lv\Psi_{\ell_2,q,\lambda_2}\ra
  \la\Psi_{\ell_2,q,\lambda_2}\rv
  v_{\beta}\lv\Psi_{\ell_1,q,\lambda_1}\ra
  \label{eq:Kubo}
  ,
\end{multline}
where $S$ is the area of the ribbon, $f(x)$ the Fermi distribution
function, and $\la \Psi_{\ell,q,\lambda}\rv v_{\alpha}\lv
\Psi_{\ell',q,\lambda'}\ra$ is the matrix element of the $\alpha$
component of the velocity operator
\cite{PhysRevB.78.085432}. Since the energy scale is determined by
$t$, let us introduce a dimensionless energy parameter $\Omega =
\hbar\omega/t$.

Translation invariance along the longitudinal direction dictates that 
the matrix elements of the velocity $v_x$ are diagonal in $q$ and
$\ell$, leading to $\sigma_{xx}$ of the form
\begin{equation}
  \Re\frac{\sigma_{xx}}{\sigma_0} = \mathcal{N}_x
  \sum_{\ell_0}\delta f_{q_0,\ell_0} M^2_x(q_0,\ell_0)
  ,
  \label{eq:sigxx}
\end{equation}
where $\delta f_{q_0,\ell_0} = f(E_{\ell_0,q_0,-})-
f(E_{\ell_0,q_0,+})$, $\mathcal{N}_x = 4/3\sqrt{3}(N-1)$, and $q_0$ is
given by
\begin{equation}
  q_0 =
  2\arccos\frac{(\Omega/2)^2-1-4\cos^2k_{\ell_0}}{4\cos k_{\ell_0}}
  .
  \label{eq:q0xx}
\end{equation}
The sum in \Eqref{eq:sigxx}  is restricted to those values of $\ell_0$
such that $q_0 \in \mathbb{R}$. Finally, $M^2_x(q_0,\ell_0)$ reads
\begin{equation}
  M^2_x(q_0,\ell_0) =
    \frac{
      \bigl[\cos\theta_{\ell_0,q_0} \!\!-\!
      \cos(\theta_{\ell_0,q_0} \!\!\! - \!q_0/2) \cos k_{\ell_0}
\bigr]^2
    }{
    \sin(q_0/2)\cos k_{\ell_0}
    }
  \label{eq:Mx}
  .
\end{equation}
Only \emph{inter}-band transitions (from the
sub-bands with $\lambda=-1$ to $\lambda=+1$) contribute to
$\sigma_{xx}$.

The analytical expression for $\sigma_{yy}$ is slightly more
cumbersome than the previous one, due to the absence of translation
invariance along that direction. As a result, (i) the matrix elements
of the operator $v_y$ are non-diagonal in the sub-band index $\ell$,
and (ii) there are both \emph{intra}-band ($\lambda=\lambda'$) and
\emph{inter}-band ($\lambda\ne\lambda'$) contributions to the
transverse conductivity. The calculation is, nevertheless,
straightforward, yielding
\begin{equation}
  \Re\frac{\sigma_{yy}}{\sigma_0} = \mathcal{N}_y
  \sum_{\ell_1,\ell_2}\sum_{\lambda,\lambda'}
  \mathcal{P}_{\ell_1,\ell_2} \,
  \delta f_{q_0,\ell_1,\ell_2}^{\lambda,\lambda'} \,
  M^2_y(q_0,\ell_1,\ell_2)
  \label{eq:sigyy}
  ,
\end{equation}
where $\mathcal{N}_y=4/\sqrt{3}(N+1)(N^2-1)$,
$\delta
f_{q_0,\ell_1,\ell_2}^{\lambda,\lambda'}=n_F(E_{\ell_1,q_0,\lambda })-
n_F(E_{\ell_2,q_0,\lambda'})$, and
$\mathcal{P}_{\ell_1,\ell_2}=1-(-1)^{\ell_1+\ell_2}$.
This latter factor entails the selection rule for transitions among
sub-bands $\ell_1+\ell_2=$~odd. The last factor is
\begin{multline}
  M^2_y(q_0,\ell_1,\ell_2) =
    \frac{
    \sin^2k_{\ell_1}\sin^2k_{\ell_2}
    }{
    \sin^2[(k_{\ell_1}+k_{\ell_2})/2]\sin^2[(k_{\ell_1}-k_{\ell_2})/2]
    }\\
  \times
    \frac{\epsilon_{\ell_1,q_0}\epsilon_{\ell_2,q_0}
    \vert\sin(q_0/2)\vert^{-1} (\hbar\omega)^{-1} }
    {\left\vert
    \cos k_{\ell_1}\epsilon_{\ell_2,q_0}+\lambda\lambda'
    \cos k_{\ell_2}\epsilon_{\ell_1,q_0}
    \right\vert}
  \times
    \mathcal{C}_{q_0,\ell_1,\ell_2}
  \label{eq:My}
  ,
\end{multline}
where $\mathcal{C}_{q_0,\ell_1,\ell_2} =
1 + \lambda\lambda'\cos(\theta_{\ell_1,q_0}+\theta_{
\ell_2,q_0}-q_0)$, and 
\begin{equation}
  q_0 =
  2 \arccos \frac{(a_2-a_1)Q_b+\Omega^2(b_1+b_2)\pm Q_c}{(b_1-b_2)^2}
  ,
  \label{eq:q0_yy}
\end{equation}
with $Q_c = 2\sqrt{\Omega^4 b_1 b_2+\Omega^2 Q_b Q_a}$,
$Q_b = b_1 - b_2$, $Q_a = b_1 a_2 - b_2 a_1$,
$a_i = 1 + 4\cos^2k_{\ell_i}$ and $b_i = 4\cos k_{\ell_i}$.
The sum in \Eqref{eq:sigyy} is also restricted to those
$\ell_1,\ell_2$ such that $q_0\in\mathbb{R}$, and to
$\lambda\le\lambda'$ (photon absorption only).

The expressions in Eqs. (\ref{eq:sigxx}) and (\ref{eq:sigyy}) are our
central result, and from them follow all the averages and other
physical quantities described and analyzed below.

\begin{figure}
  \centering
  \includegraphics*[width=0.5\textwidth]{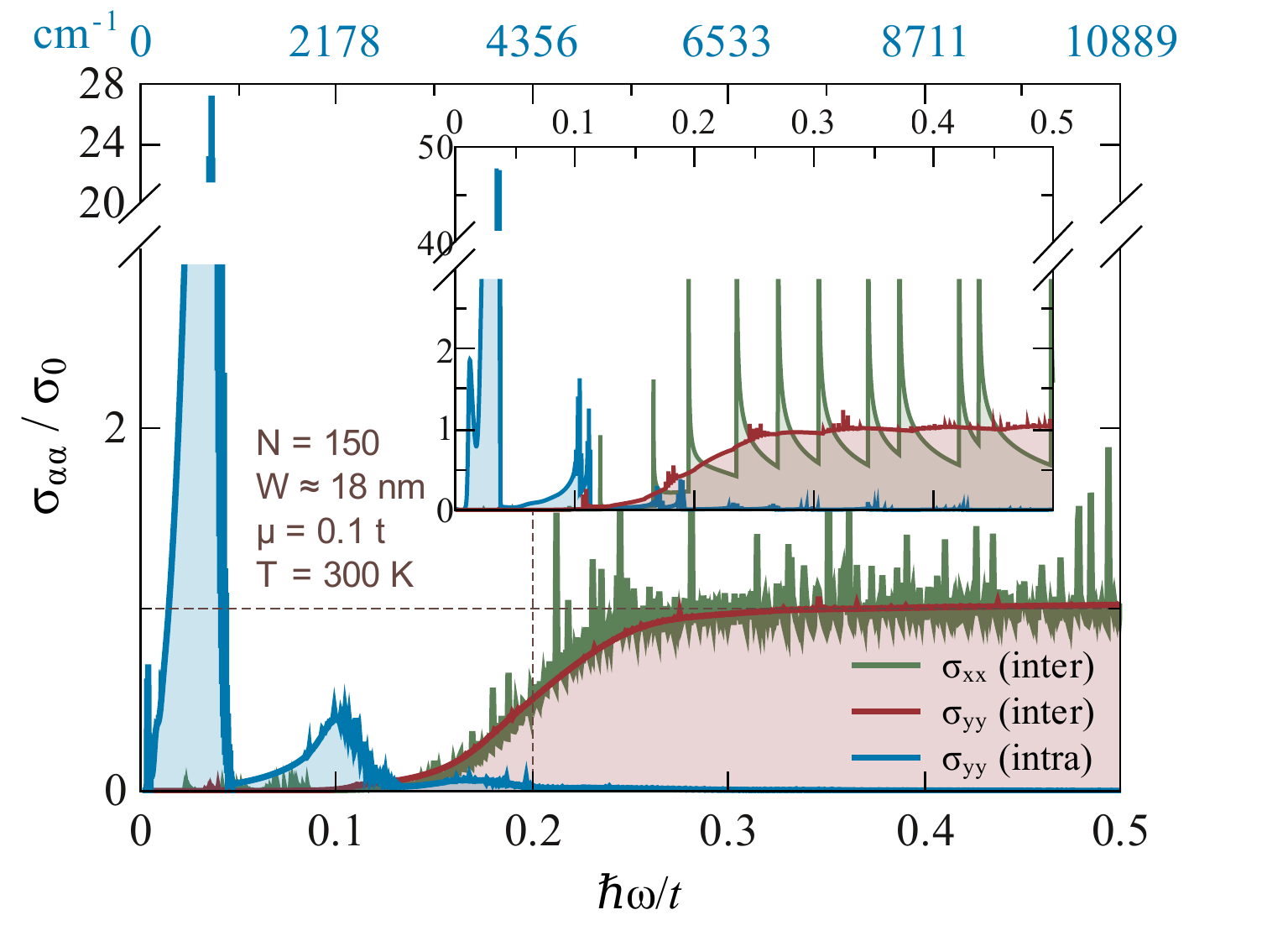}%
  \caption{
    The three non-zero contributions for
    $\avg{\sigma_{\alpha\alpha}(\omega)}$ discussed in the text,
    showing a very strong anisotropy in the infrared.
    In this example the optical conductivities are calculated for
    an ensemble of ribbons having
    $ \avg{N} = 150$ ($\simeq 18.5$\,nm),
    $\sqrt{\la N^2 -\avg{N}^2\ra} = 10$ ($\simeq 1.2$\,nm).
    We further used $T=300K$ and $\mu = 0.1$ ($\simeq 0.3$\,eV).
    The \inter
    contributions essentially follow the bulk 2D behavior, with the
    expected temperature-broadened step onset at $\hbar\omega =
    2\mu$. In contrast, the \intra contribution for the transverse
    conductivity ($\sigma_{yy}$) is strongly peaked at low  energies.
    Also note that the vertical axis is \emph{truncated} for clarity,
    and that $\sigma_{yy}$ peaks at nearly $28\sigma_0$ for this
    ensemble.
    The inset shows the same three quantities, but for a single ribbon
    of $N=150$, rather than the ensemble.
  }
  \label{fig:Sigma}
\end{figure}

%
\section*{Anisotropic Optical Absorption}

Lateral confinement, reduces the energy spectrum of GNRs to a set of
subbands, each reflecting the dispersion of an effective 1D mode
$\ell$ ($\ell=1,2,\ldots,N$), propagating longitudinally with momentum
$q$: \mbox{$E_{\ell,q,\lambda}=\lambda t\,\epsilon_{\ell,q}$}, where
$\lambda=\pm1$, defines the valence and conduction subbands,
\begin{equation}
  \epsilon_{\ell,q} = \sqrt{1+4\cos k_\ell\cos(q/2)+4\cos^2k_\ell}
  \label{eq:Dispersion}
  ,
\end{equation}
and $k_\ell$ is transverse quantized momentum: $k_\ell=\pi\ell/(N+1)$.
Consequently, the density of states is dominated by Van Hove
singularities (VHS) that develop at $q=0$ for each subband
\cite{Fujita:1996,Katsunori,Wakabayashi:1999}. Such sharp spectral
features translate into strong optical absorption for ideal GNRs, but
are readily smoothed out by edge or bulk disorder and/or temperature
in real systems \cite{EndNote-4}. Our ensemble averaging has the same
effect.

In \Fref{fig:Sigma} we show the averages $\avg{\sigma_{xx}}$ and
$\avg{\sigma_{yy}}$ for an ensemble with $\avg{N}=150$, and finite
chemical potential: $\mu=0.1$. This particular value of chemical
potential was chosen to allow a clear distinction between the \inter
and \intra contributions to the conductivity, so as to
better illustrate the main features of the absorption spectrum. As a
consequence of time reversal
symmetry, only the diagonal components of $\sigma_{\alpha\beta}$ in
the coordinate system of \Fref{fig:Illustration} are non-zero.
Translation invariance along the longitudinal ($x$) direction implies
that only \inter transitions contribute
to $\sigma_{xx}(\omega)$, as derived explicitly above. Consequently,
$\avg{\sigma_{xx}(\omega)}$ reproduces the bulk 2D behavior, as is
clearly seen in \Fref{fig:Sigma}. For the analysis of the transverse
conductivity, $\avg{\sigma_{yy}(\omega)}$, it is convenient to isolate
the \emph{inter-} and \emph{intra-}band contributions:
$\avg{\sigma_{yy}(\omega)} = \avg{\sigma_{yy}^\text{inter}(\omega)} +
\avg{\sigma_{yy}^\text{intra}(\omega)}$ 
(the latter is allowed since along the transverse direction the
electron scatters off the ribbon edges). Whereas
$\avg{\sigma_{yy}^\text{inter}}$ featurelessly
follows $\avg{\sigma_{xx}}$ (and hence the bulk 2D behavior), its
\intra counterpart displays a rather strong feature at low energies
which, for this specific example, nearly reaches 30 times the
universal value $\sigma_0$.

\begin{figure}
  \subfigure[][]{%
    \includegraphics*[width=0.18\textwidth]{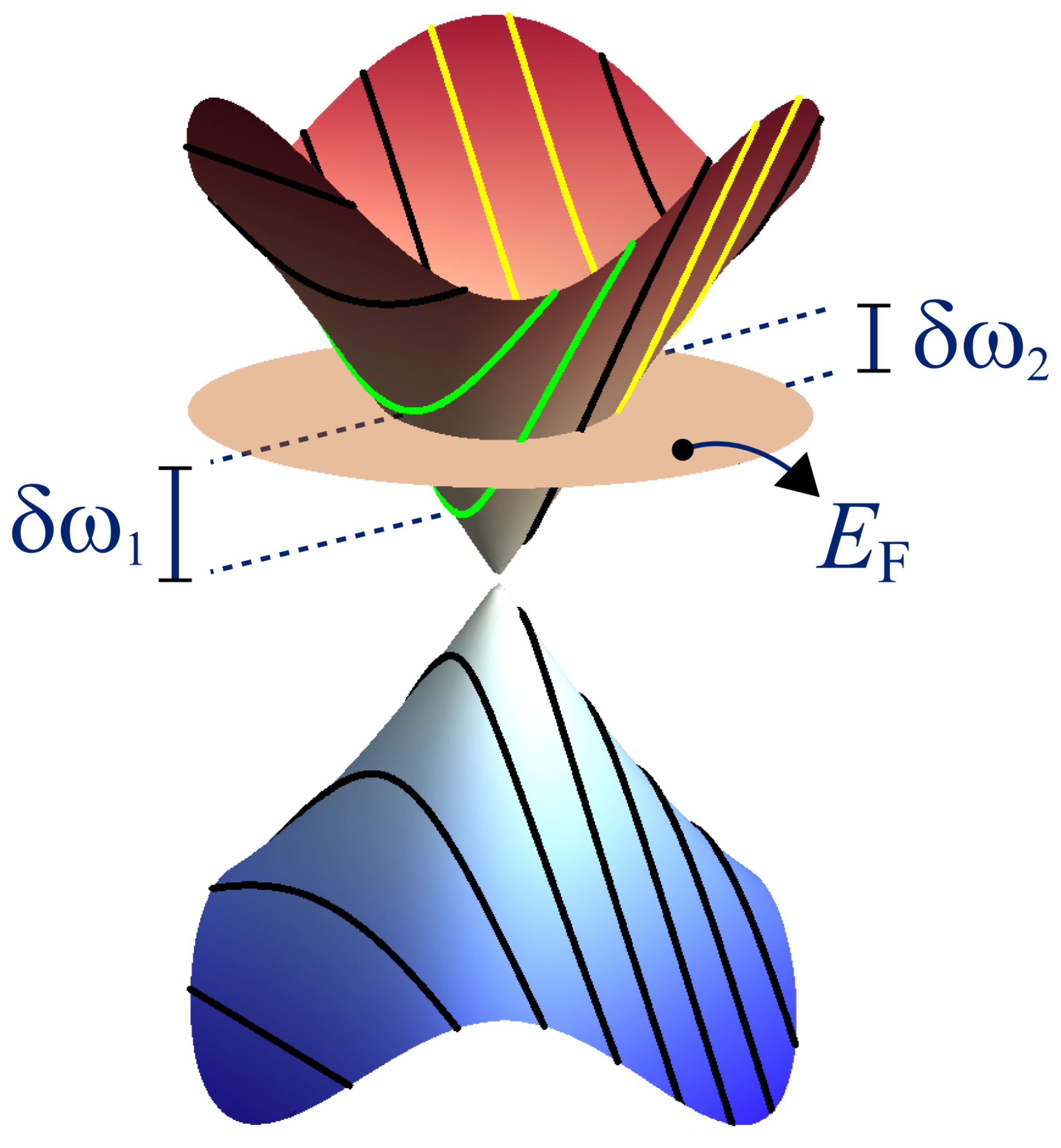}%
    \label{fig:Spectrum-1}%
  }%
  \subfigure[][]{%
    \includegraphics*[width=0.3\textwidth]{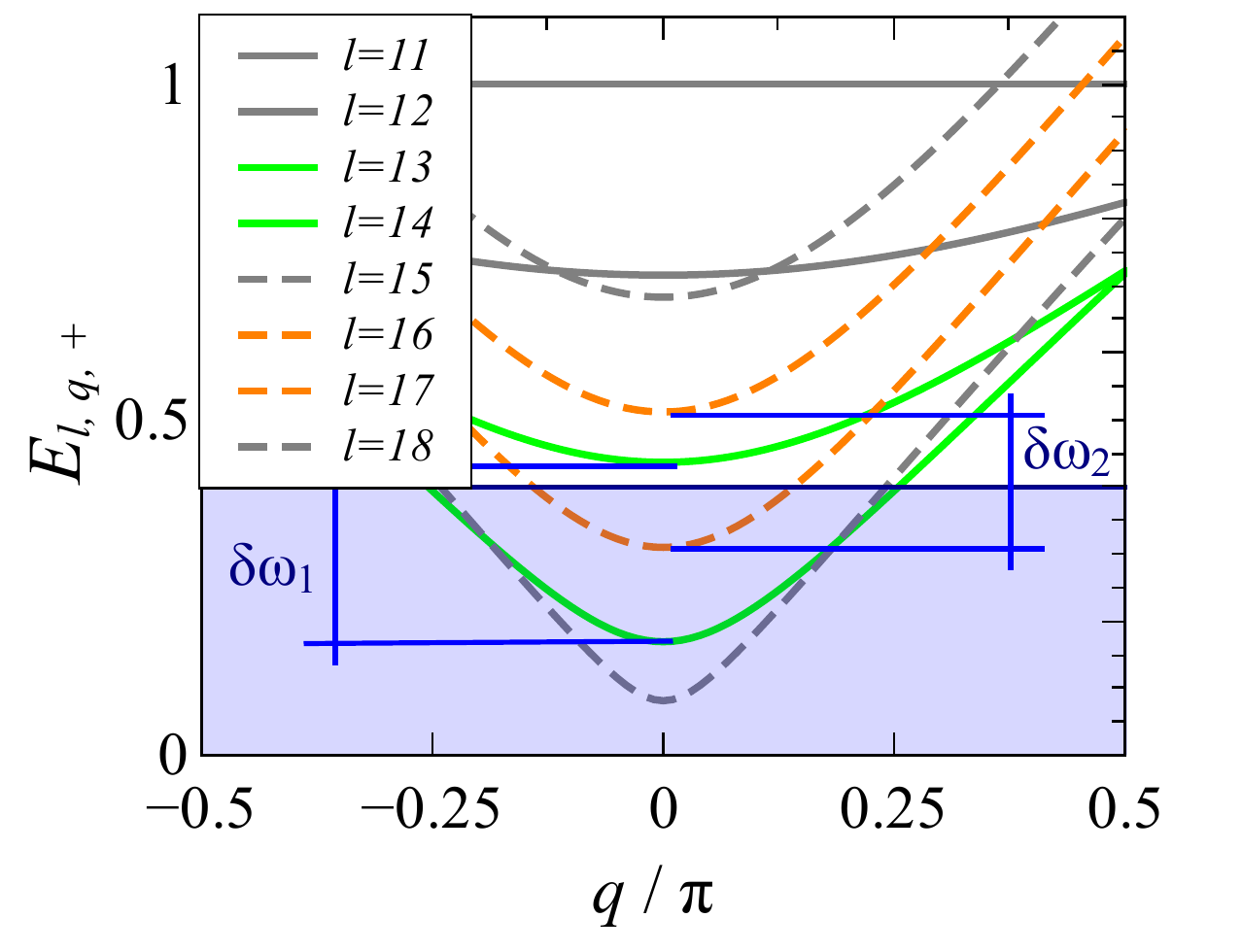}%
    \label{fig:Spectrum-2}%
  }
  \caption{
    \subref{fig:Spectrum-1}
    Illustration of the transverse spectrum quantization and highlight
    of the two most significant \intra
    transitions $\delta\omega_{1,2}$,
    which appear around the Dirac point ($N=21$, $E_F=0.4t$).
    \subref{fig:Spectrum-2}
    The projection of the subbands $E_{\ell,q,+}$ shown in
    \subref{fig:Spectrum-1}, as a function of the 1D momentum $q$, and
    for $l=11,\dots,18$ (see text).
  }
  \label{fig:Spectrum}
\end{figure}

\begin{figure}
  \centering
  \includegraphics*[width=0.5\textwidth]{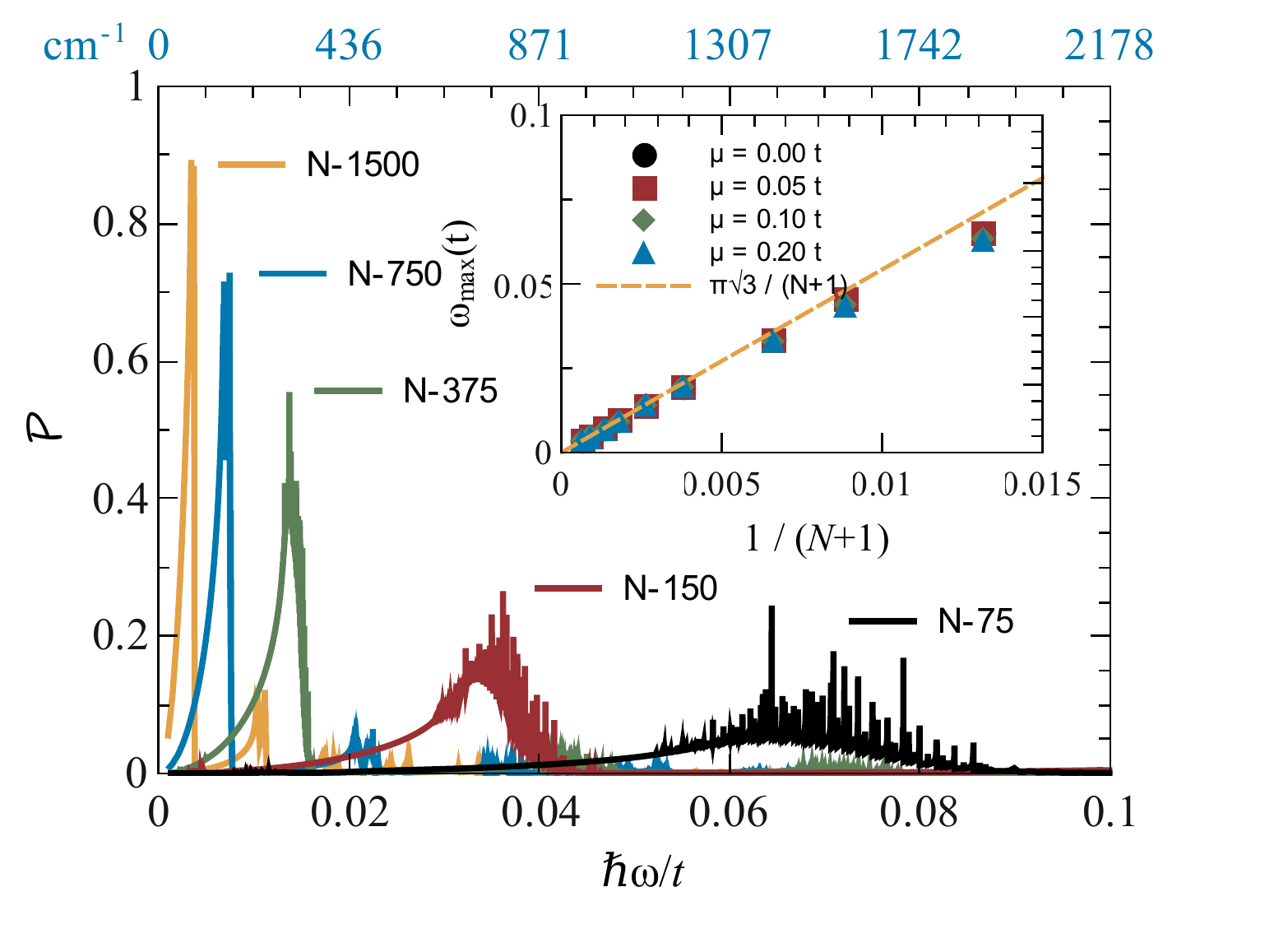}%
  \caption{
    The degree of polarization $\cal P(\omega)$ in the low energy
    region, for
    ribbons of different average width (in unit cells) $\avg{N}$ and
    $\mu=0.1$, $T=300K$
    (for reference,
    $\avg{N}=\{75,\,150,\,375,\,750,\,1500\}\;\Leftrightarrow
    \avg{W}=\{9,\,18,\,46,\,92,\,184\}$\,nm).
    The inset shows the position of the most
    prominent peak in $\avg{\sigma_{yy}^\text{intra}}$ as a function
    of $\avg{N}$ and $\mu$. The $\mu$ dependence is expectedly
    weak, while the peak position is seen to follow the analytical
    form described in the text.
  }
  \label{fig:P-vs-N}
\end{figure}

Some aspects of \Fref{fig:Sigma} are worth underlying. Firstly, it is
evident that, despite averaging to the same step-wise
$\omega$-dependence, $\avg{\sigma_{yy}^\text{inter}(\omega)}$ is much
smoother than $\avg{\sigma_{xx}^\text{inter}(\omega)}$, even though
the averages are over the same ensemble. This can be traced to the
fact that, for each individual ribbon, only $N$ symmetric transitions
($-E \to +E$) contribute to $\sigma_{xx}^\text{inter}(\omega)$,
whereas $\sigma_{yy}^\text{inter}(\omega)$ includes
$\mathcal{O}(N^2)$ transitions among almost all pairs of subbands.
Consequently, the latter has many more absorption singularities, but
much weaker, by conservation of spectral weight (this is explicitly
shown in the inset of \Fref{fig:Sigma}). The averaging is thus more
efficient in washing out the structure of VHSs in
$\avg{\sigma_{yy}^\text{inter}(\omega)}$.
Secondly, the low-energy peak in
$\avg{\sigma_{yy}^\text{intra}(\omega)}$ can be already identified
from a single ribbon (inset). Its origin is simple to understand with
reference to \Fref{fig:Spectrum}. Since the band structure consists of
a set of discrete subbands, the chemical potential will always be
straddled by two of them at $q=0$, such that
$E_{\ell,q=0,\lambda}<\mu<E_{\ell+1,q=0,\lambda}$. Given that
transitions $\ell\to\ell+1$ are allowed in $\sigma_{yy}^\text{intra}$,
one expects an absorption peak at $\hbar\omega \approx
|E_{\ell,q,\lambda}-E_{\ell+1,q,\lambda}|$. Moreover, as per
\Eqref{eq:My} the matrix element decays rapidly with the difference in
band index, so that the transitions between the two bands closest to
$\mu$ completely dominate $\sigma_{yy}^\text{intra}$. From
\Fref{fig:Spectrum} it is clear that there are always two pairs of
such bands, whose energy difference at $q=0$ is
$\hbar\,\delta\omega_{1,2}\approx\pi \sqrt{3-\mu^2\pm2\mu}/(N+1)$.
Since we are interested in situations where $\mu\ll 1$, the \intra
peaks are solely determined by the ribbon geometry:
$\hbar\omega_\text{max}\approx\pi \sqrt{3}/(N+1)$. This can be
confirmed in the inset of \Fref{fig:P-vs-N} for ensembles with
different $\avg{N}$, and introduces an element of
\emph{predictability} and \emph{tunability} with respect to the
frequency band where the optical absorption is highly enhanced. In
other words, given the frequency of operation desired for a given
application, one can select the appropriate average ribbon width that
yields the strongest optical anisotropy at that target frequency.

\begin{figure}
  \hspace*{-2em}
  \subfigure[][]{%
    \includegraphics*[width=0.225\textwidth]{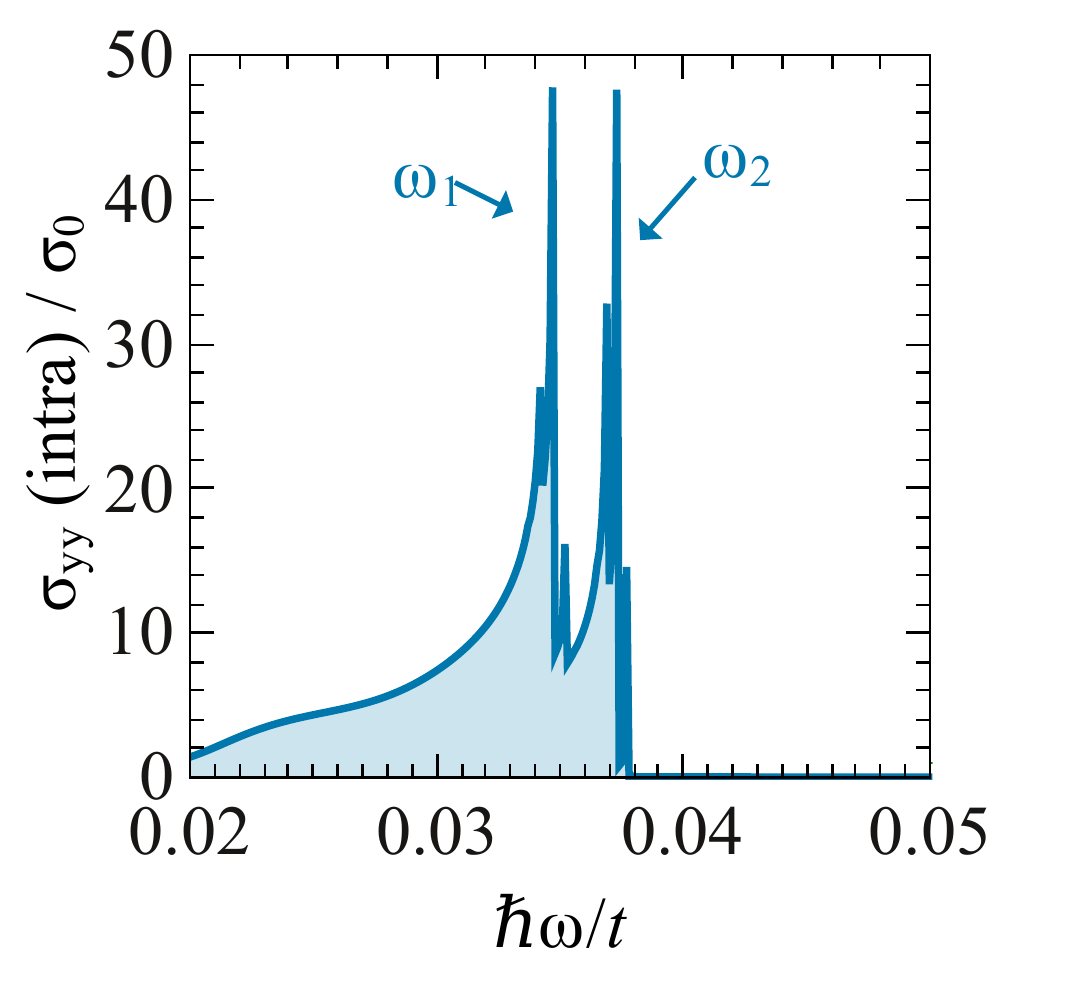}%
    \label{fig:DoublePeak-1}%
  }%
  \subfigure[][]{%
    \includegraphics*[width=0.25\textwidth]{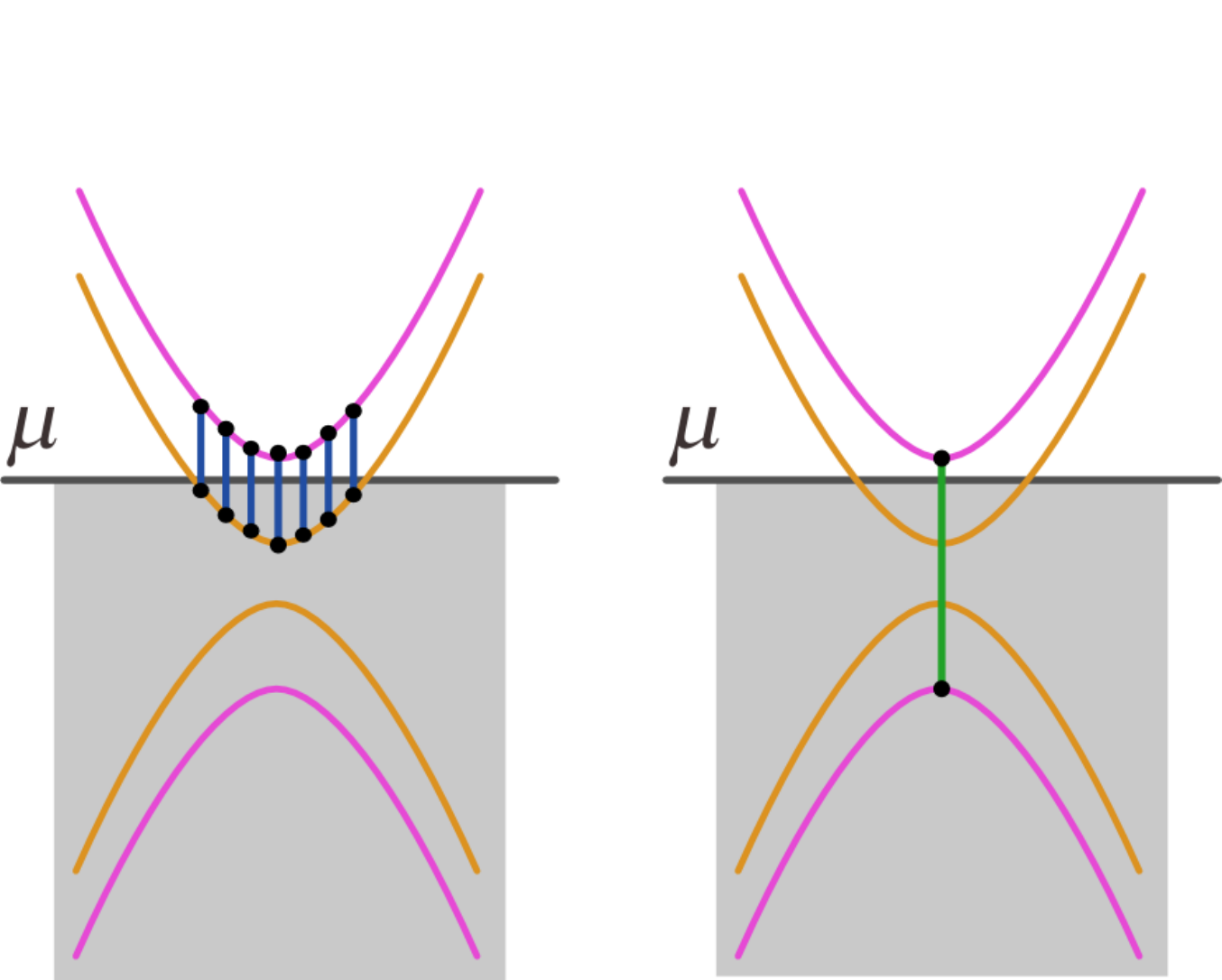}%
    \label{fig:DoublePeak-2}%
  }
  \caption{
    \subref{fig:DoublePeak-1}
    A close-up of the most prominent feature
    in $\sigma_{yy}^\text{intra}$ for the single ribbon whose
    absorption spectrum was shown in the inset of \Fref{fig:Sigma}.
    On this close-up the double peak structure discussed in the text
    is clearly visible.
    \subref{fig:DoublePeak-2}
    An illustration of the different nature of the processes
    contributing to $\sigma_{yy}^\text{intra}$ and
    $\sigma_{xx}^\text{inter}$. In the former, the resonant
    transitions occur between quasi-parallel sub-bands, whereas in
    the latter the resonant condition is only strictly verified at
    the van Hove point.
  }
  \label{fig:DoublePeak}
\end{figure}

Another relevant detail to notice is that, as seen in
\Fref{fig:Sigma}, the absorption peak in $\sigma_{yy}^\text{intra}$ is
much more resilient to the ensemble averaging (or level broadening)
than all the other transitions coming from \inter processes: the
averaging readily washes out the VHS features, but leaves the peak in
$\avg{\sigma_{yy}^\text{intra}}$ quite well defined and intense. In
the case shown in \Fref{fig:Sigma}, $\avg{\sigma_{yy}^\text{intra}}$
peaks at a few dozen times the value of the longitudinal
$\avg{\sigma_{xx}}$ The reason for this is very simple to understand
qualitatively, and is twofold. On the one hand, since there are
always two resonant conditions very close in frequency
(for example, $\delta\omega_1$ and $\delta\omega_2$ in
\Fref{fig:Spectrum}), the shape of the feature in
$\sigma_{yy}^\text{intra}$ has a double peak structure. To show this
explicitly, in \Fref{fig:DoublePeak} we present a close-up of
$\sigma_{yy}^\text{intra}$ for the single ribbon with $N=150$
previously shown in the inset of \Fref{fig:Sigma}: the double peak
structure is self-evident.
In addition to that,
the transition processes contributing to $\sigma_{yy}^\text{intra}$
are quite different from the ones that contribute to
$\sigma_{xx}$, or $\sigma_{yy}^\text{inter}$. In a single independent
ribbon, the longitudinal conductivity is dominated by \inter
transitions among sub-bands which have an inverted dispersion with
respect to each other (see \Fref{fig:DoublePeak}(b) for an
illustration). Consequently the
resonant condition occurs only at the van Hove point, leading to the
very sharp van Hove absorption peaks in $\sigma_{xx}$ that we see in
the inset of \Fref{fig:Sigma}. In
contrast, the processes contributing the most to
$\sigma_{yy}^\text{intra}$ involve transitions among \emph{nearly
parallel} sub-bands [\Fref{fig:DoublePeak}(b)], thus allowing a finite
density of momentum states to contribute to the resonance, and
implying a larger joint density of states. This makes the absorption
feature in $\sigma_{yy}^\text{intra}$ broader than the van Hove-type
peaks associated with $\sigma_{xx}$. The consequence of this is
that, when one considers the ensemble averaging, the sharp van Hove
peaks in the longitudinal conductivity will be slightly displaced with
the changing $N$ within the ensemble, and are rapidly washed out. The
double-peak structure, combined with the broader
\emph{parallel}-dominated absorption, \emph{protects} the transverse
absorption peak with respect to the level broadening, thereby
resulting in an absorption feature that is much more robust.

To assess the polarizing efficiency of a single graphene ribbon we
calculate the optical transmission amplitude, which is the ratio of
the electric field amplitudes of the incoming and transmitted fields:
$t_\alpha(\omega) =  E^{(t)}_\alpha / E^{(i)}_\alpha $,
($\alpha=x,y$). For radiation impinging normally upon an ensemble of
GNRs separating medium 1 and medium 2 (\Fref{fig:Illustration}), the
transmission amplitude reads explicitly
\begin{equation}
  t_\alpha(\omega) = \frac{2\,Z^{(2)}}
    {Z^{(1)}+Z^{(2)} [1 + Z^{(1)}\avg{\sigma_{\alpha\alpha}(\omega)} ]
}
  \label{eq:Transmission}
  ,
\end{equation}
where $Z=\sqrt{\mu_0\mu/\epsilon_0\epsilon}$ is the impedance of each
medium. This result is obtained in the conventional way, by assuming
that the system of graphene ribbons is a metallic sheet of zero
thickness, and imposing the boundary conditions of the electromagnetic
field at the interface. Knowledge of $t_\alpha(\omega)$ allows for the
calculation of the \emph{degree of polarization} (DP,
$\mathcal{P}(\omega)$), or the rotation of the
plane of linear polarization ($\theta = \theta_f-\theta_i$):
\begin{equation}
  \mathcal{P}(\omega) = \frac{|t_x|^2 - |t_y|^2}{|t_x|^2 + |t_y|^2}
   ,\qquad
   \tan\theta_f = \frac{t_y(\omega)}{t_x(\omega)} \tan\theta_i
   \label{eq:Polarizability}
  ,
\end{equation}
This definition is useful for unpolarized incoming light where
$\mathcal{P} = \pm 1$ reflects full polarization of the incoming wave.
For an already polarized incoming wave, the second equation shows
that the effect naturally depends on the orientation of the incoming
polarization with respect to the ribbon principal directions.
With $\mathcal{P}(\omega)$ we can immediately identify the degree of
dichroism by how close $|\mathcal{P}(\omega)|$ is to unity (i.e. how
close to an ideal polarizer are we).

In \Fref{fig:P-vs-N} we plot $\mathcal{P}(\omega)$ for different
ribbon widths. It can be clearly seen that, DP in excess of 50\%
can be achieved already with ribbons $45$\,nm wide. We underline that
\emph{this is the degree of polarization produced by an atomically
thin ensemble of ribbons}, which makes the magnitude of the effect
even more striking! Even though the transparency of infinite 2D
graphene is as large as 97.7\%, the confinement-induced anisotropy can
be so large as to almost completely suppressing one of the field
projections.
The same figure also confirms that the optimum DP
is achieved at a width-dependent frequency $\omega_\text{max}$ which,
as discussed above, has a simple form (inset of \Fref{fig:P-vs-N}).
However it is also clear that this tunability is at the expense of the
absolute amount of DP (narrower ribbons $\to$
larger $\omega_\text{max}$ $\to$ smaller
$\mathcal{P}(\omega_\text{max})$). Nevertheless, it has been
experimentally confirmed that the optical absorption of $N$-layer
graphene is simply proportional to $N$, from the bilayer
 to graphite\cite{EndNote-3} for most of the low energy range
\cite{Kuzmenko:2008,Science_Nair,Heinz:2010}. This
means that the effect reported here can be significantly magnified by
using few-layer graphene ribbons, or simply
superimposing a few independent layers onto each other.

\begin{figure}
  \includegraphics*[width=0.5\textwidth]{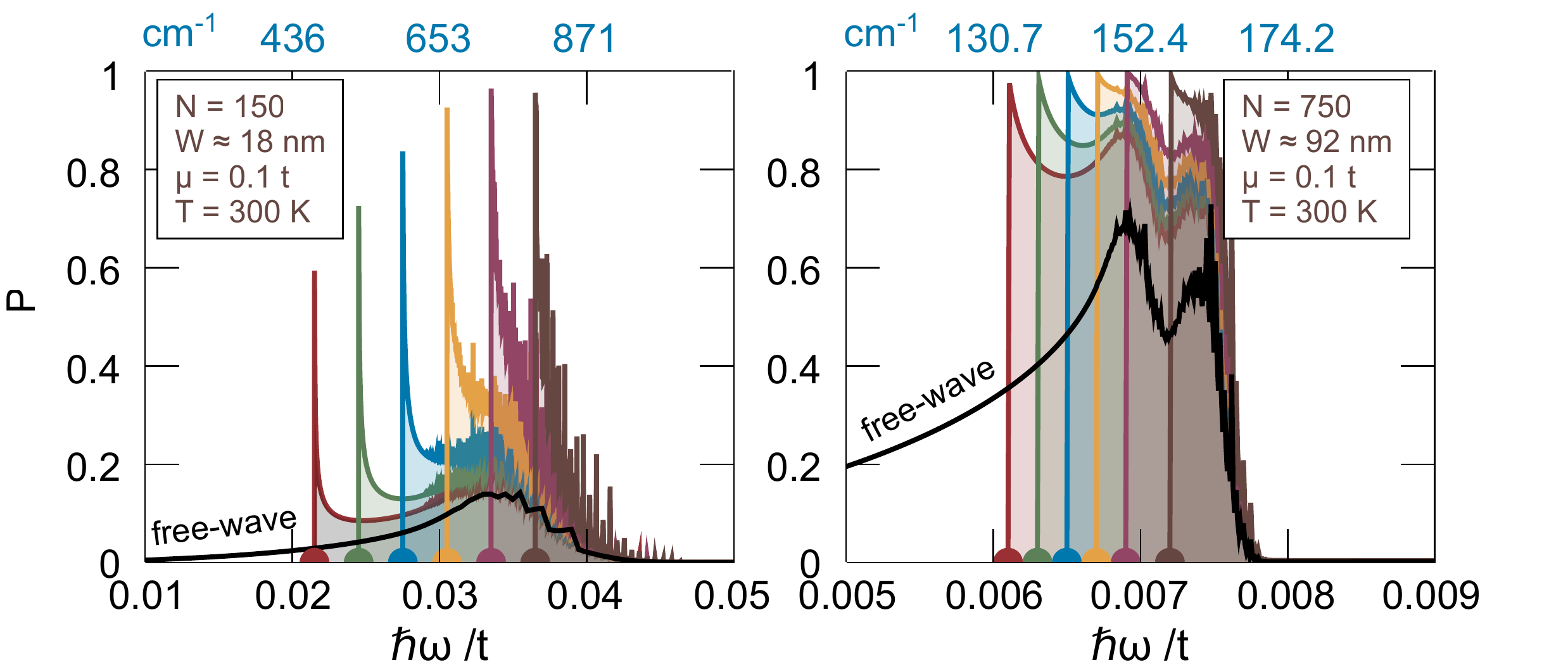}
  \caption{
    The effect of a metallic waveguide of square cross-section
    in the degree of polarization
    $\mathcal{P}(\omega)$ for two ensembles of ribbons
    ($\avg{N}=150,\,750$). Each panel shows $\mathcal{P}(\omega)$
    for an incoming wave made of a combination \cite{EndNote-1} 
    of the two lowest degenerate modes TE$_{10,01}$, in vacuum (black)
    and in waveguides (colors) with different geometries, \emph{i.e.}
    different cut-off frequency $\omega_{10}$. Each $\omega_{10}$ is
    marked by a dot at the corresponding w in the horizontal axis
    and a unique color.
  }
  \label{fig:P-cavity}
\end{figure}

In addition, the form of \Eqref{eq:Transmission} given in terms of the
impedance of the media suggests that additional parameter freedom can
be achieved if the wave propagates inside a metallic waveguide. As is
well known, electromagnetic propagation in waveguides is restricted to
normal TEM, TM or TE modes. Each of the latter two has a
characteristic dispersion that is different from the free-space
relation $\omega = c k / n$. For the purpose of analyzing transmission
and reflection amplitudes in a situation as depicted in
\Fref{fig:Illustration}, the effect of the waveguide can be absorbed
in a renormalized and frequency-dependent impedance, $Z(\omega)$. For
example, the mode TE$_{mn}$ has a characteristic impedance
\cite{Jackson} $Z_{mn}(\omega)=Z \omega / \sqrt{\omega^2 -
\omega_{mn}^2}$, where $\omega_{mn}^2 = (c^2\pi^2 / \mu\epsilon)
[(m/a)^2+(n/b)^2]$. Hence each mode can only propagate if $\omega$ is
beyond the mode cut-off frequency $\omega_{mn}$, and this is
frequently used to select/restrict the propagating modes by adapting
the geometry of the waveguide. In our example we could take a square
cross section ($a=b$), in which case the two degenerate modes
TE$_{10}$ and TE$_{01}$ can be combined into an arbitrary incoming
plane polarization \cite{EndNote-1}. In that case, if $\omega_{10} <
\omega < \omega_{11}$, only the modes TE$_{10,01}$ propagate in the
waveguide, and $Z_{10}(\omega)=Z \omega / \sqrt{\omega^2 -
\omega_{10}^2}$. The cavity setup is interesting and useful for two
reasons, which can be understood by inspection of \Fref{fig:P-cavity}:
(i) on one hand, by tuning the cavity dimensions so that
$\omega_{10}\lesssim\omega_\text{max}(N)$ one can precisely cut-off
the DP below $\omega_{10}$, creating a well
defined band of frequencies where the system displays high DP; (ii) on
the other hand, since $Z_{10}(\omega) > Z$ (and,
in particular $Z_{10}(\omega\gtrsim\omega_{10}) \gg Z$), the cavity
highly magnifies the DP, even for a monolayer
system. Taking as illustration the ribbon ensemble with $\avg{N} =
750$ shown in \Fref{fig:P-cavity}, proper tuning of the cut-off
frequency can introduce a clear and well defined band filter for
$\mathcal{P}(\omega)$, while simultaneously amplifying the magnitude
of $\mathcal{P}(\omega)$ in comparison with the value for a free wave.
($\mathcal{P}(\omega)$ climbs beyond 80\% in the entire frequency
window). Lastly, this enhancement of the impedance can also
make $\mathcal{P}(\omega)$ more \emph{step-like} within the strongly
amplified regime, rather than \emph{peak-like}, as implied by the
right panel of \Fref{fig:P-cavity}.

%
\section*{Discussion}

The optical absorption of a ribbon is seen here to be highly
anisotropic on account of the new \intra channel made possible by the
finite transverse direction, and the resulting electron scattering at
the ribbon edges. 
Recent experiments do show that the transmission spectrum of graphene
ribbon arrays is rather different for light polarized parallel
and perpendicularly to the ribbon length, with the latter dominated by
a plasmon absorption resonance at $\sim 3$\,THz \cite{Ju2011}.
However, these experiments pertain to ribbons much wider
($\gtrsim1\,\mu\text{m}$) than the ones envisaged here ($\lesssim
50\,\text{nm}$), such that their spectrum is effectively continuous.
Naturally, in the limit of wide ribbons
($N\to\infty$), the peak of $\sigma_{yy}$ in \Fref{fig:Sigma}
simultaneously narrows and moves towards $\omega=0$, where it becomes
the Drude singularity that we expect for an infinite and disorder-free
system. Indeed, the easiest way to understand the sharp feature of
$\sigma_{yy}$ at low energies is to see it as a usual Drude peak that
has been shifted to finite $\omega$ by making the system finite along
the transverse direction, thus allowing \intra transitions of finite
frequency.

The issue of how to actually manufacture a grid of narrow GNRs with
consistent and predictable width has been addressed earlier. It can be
achieved by means of high precision patterning using a He-ion beam
microscope in lithography mode \cite{Lemme:2009}, or more standard
etch masks able to cut down to the 10\,nm scale \cite{Bai:2009}. An
alternative to cutting ribbons out of graphene sheets is the recently
developed technique of unzipping carbon nanotubes (CNTs)
\cite{Kosynkin:2009,Hongjie:2010,Crommie:2011}. Nowadays it is
possible to produce batches of CNTs with similar radius
\cite{HongjieDai:2007}, and so this would allow for the production of
high quality ribbons without edge disorder. Another alternative, that
completely bypasses patterning, consists in inducing effective
nanoribbons by engineering a periodic distribution of strain in a
bulk graphene sheet, such that the strain-induced confinement
mimics the ribbon quantization features \cite{vitorbreak}.

As always in the context of GNRs, the role of disorder needs to be
addressed, and perhaps electron-electron interactions as well
\cite{PRL10}. It is known that disorder can affect and even destroy
many intrinsic features, such as the edge modes in zig-zag (ZZ) GNRs
\cite{Fujita:1996}, the spontaneous spin polarization expected for
ideal ZZ ribbons \cite{Wakabayashi:1999,Son:2006}, the width scaling
of the gap \cite{Stampfer:2009,GoldhaberGordon:2010}, or their
conductance \cite{Mucciolo:2009}. In our case, disorder can modify the
intrinsic optical anisotropy in different ways, depending on the
causes: (i) inhomogeneities of the free  carrier density
caused by various external effects (e.g., substrate inhomogeneities,
asdorbates, charged impurities); (ii) spatial fluctuations of the site
energy and hopping parameters leading to broadening of mini-bands and
carrier scattering, which in turn broadens and shifts the \intra
absorption peaks; (iii) adsorbates and other impurities can introduce
spurious features in the absorption spectrum; (iv) edge disorder can
lead to localization of some electronic states \cite{Mucciolo:2009}.
Concerning (i), typical electron density fluctuations in graphene on
representative substrates, such as SiO$_2$, have been evaluated
experimentally \cite{Martin2008}, and seen
to be of the order of $\delta n_e\sim4\times10^{10}\mathrm{cm}^{-2}$
in relatively clean systems. Such effects will presumably have little
impact when the overall carrier density is between $10^{11}-10^{12}$,
which are the densities targeted in our study.
The effects of diagonal and non-diagonal disorder (ii) are
expected to be less important for narrower ribbons, simply because the
anisotropy is induced by \emph{intra}-subband absorption, and the
separation of the subbands scales as $\propto1/N$ (and so the
narrower the ribbon the less significant become local fluctuations of
the potential energy, or the hopping amplitudes).
Therefore, it is expected that the necessary anisotropy in
$\sigma(\omega)$ might be achieved in practice.
Regarding (iii), post-patterning annealing techniques have been
progressively improved, and proven quite efficient in removing such
sources of disorder \cite{Moser:2007}; alternatively,
encapsulation of graphene has been shown to
significantly reduce environmental contamination and
to reduce electronic scattering \cite{Geim:2011}.
With respect to (iv), much
depends on the fabrication technique, and the CNT unzipping method (or
perhaps the strain-engineering route) would be preferred to mitigate
edge disorder. If present in a strong degree, however, edge disorder
might bring about new effects not considered here. In particular,
experiments show that edge disorder arising from conventional
lithographic procedures leads to strong electron localization, and the
emergence of a system of effective coupled quantum dots, where
charging and interaction effects can be important
\cite{Sols:2007,Stampfer:2009}. The extent to which these
features modify the absorption spectrum is not known experimentaly
and, theoretically, a realistic approach to the problem is out of
range of a fully analytical approach, as we seek and use here. 
These effects will be addressed in future work.

Another issue to consider is the low frequency
absorption characteristic of any metal, associated with
disorder-induced \intra transitions, and accounted for by the Drude
model. In the case of graphene, the Drude conductivity is given by
\begin{equation}
  \label{Drude}
  \frac{\sigma_D}{\sigma_0}=\frac{4\left|\mu\right|}{\pi}\frac{1}{
\hbar\left(\gamma-i\omega\right) }
\end{equation}
where $\gamma$ is the Drude scattering rate. For
nanoribbons, such a term would have to be added to $\sigma_{xx}$. 
The appearance of a Drude peak at $\omega = 0$ is not
expected to drastically affect the absorption peaks discussed so
far, which occur at $\omega=\omega_\text{max}$ (finite). A similar
conclusion was drawn in recent experiments measuring
optical absorption in nanoribbons much wider than our target
widths (and so quantization effects disappear there), which show 
anisotropic absorption features dominated by plasmon absorption, which
are vastly insensitive to the Drude component \cite{Ju2011}.

\begin{figure}
  \subfigure[][]{%
    \includegraphics*[width=0.25\textwidth]{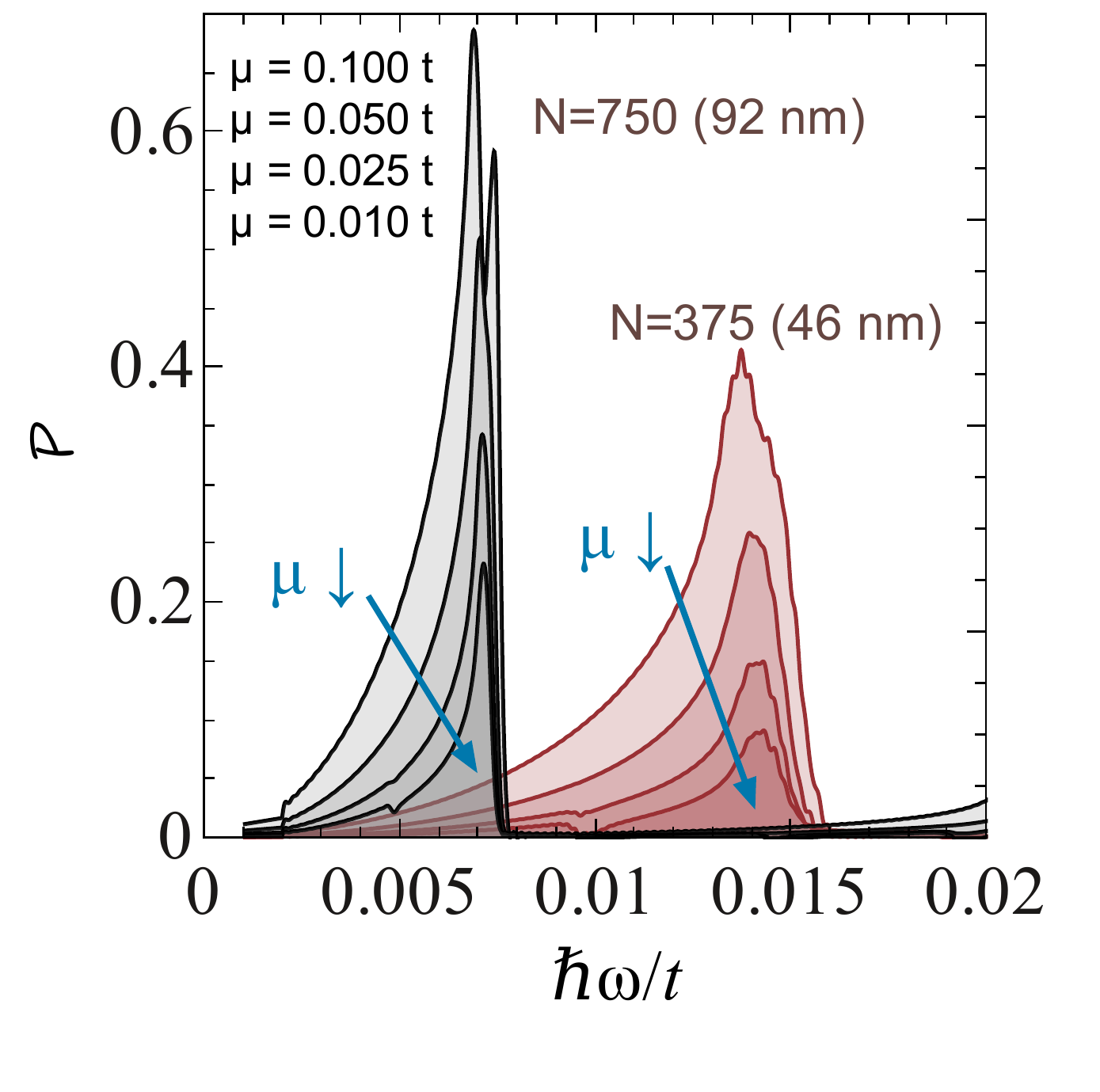}%
    \label{fig:MuDependence-1}%
  }%
  \subfigure[][]{%
    \includegraphics*[width=0.25\textwidth]{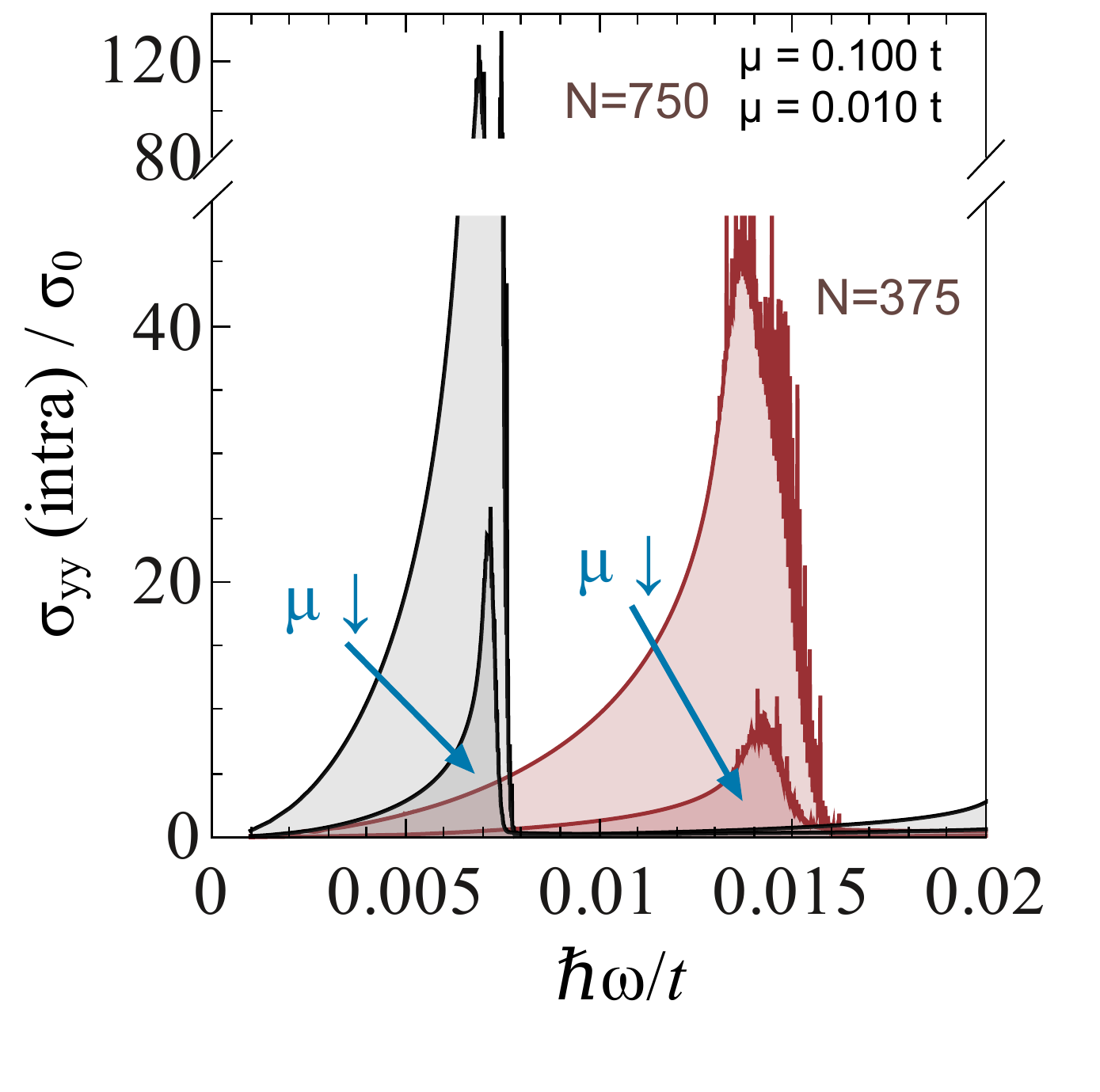}%
    \label{fig:MuDependence-2}%
  }
  \caption{
    \subref{fig:MuDependence-1}
    The degree of polarization $\cal P(\omega)$ as a function of Fermi
    energy. The behavior at low energies is shown for two ensembles of
    ribbons of average width
    $\avg{N}=750$ (92~nm) and $\avg{N}=375$ (46~nm) and, for each,
    the Fermi energy is varied from  $\mu=0.01t$ to $\mu=0.1t$.
    Lowering $\mu$ leads to the progressive decrease in the magnitude
    of the degree of polarization.
    \subref{fig:MuDependence-2}
    The same analysis but now for
    $\avg{\sigma_{yy}^\text{intra}(\omega)}$,
    and considering only the two extreme values of $\mu$. Notice how 
    the vertical scale is truncated, and that the transverse
    conductivity in the case $\avg{N}=750$ peaks at $120\sigma_0$
    for $\mu=0.1t$, and at $25\sigma_0$ for $\mu=0.01t$.
  }
  \label{fig:MuDependence}
\end{figure}

At any rate, to be more quantitative, the typical Drude scattering
rate lies in the vicinity of $100\,\text{cm}^{-1}$ ($= 0.005 t$)
\cite{Horng:2011}. Thus, in view of the results of \Fref{fig:P-vs-N},
the Drude regime should only dominate for ribbons of average width
above 184\,nm ($\avg{N}\gtrsim 1500$). Such ribbons are too wide
anyway for the sort of dimensions we are primarily interested in,
which lie around 50\,nm or below $\bigl(\avg{N}\lesssim 375\bigr)$,
and
for which we find DP in excess of 50\,\% already. In addition,
\Fref{fig:Sigma} shows that the magnitude of the peak in the
transverse conductivity easily reaches 10-20 times the value
$\sigma_0$. For wider ribbons than the one shown (184\,nm) the peak
easily surpasses a factor of 100, even after an ensemble average has
been performed (see, e.g., \Fref{fig:MuDependence}).

The Drude peak, on the other hand, has a magnitude
given by $\Re[\sigma_D(\omega =0) / \sigma_0] = 4|\mu|/(\hbar\gamma)
\approx 800\,|\mu|/t$. For $\mu=0.1t$ this means that
$\Re[\sigma_D(\omega=0) / \sigma_0] \approx 80 $. However, if we
lower the Fermi energy by a factor of 10 to $\mu=0.01t$, its
magnitude will be 10 times smaller, of course, but the change in the
transverse absorption peak is not so significant. An example of this
is shown in \Fref{fig:MuDependence}, where we show the effect of
decreasing $\mu$ (i.e. lowering the carrier density), both on the
degree of polarization, and
on $\avg{\sigma_{yy}^\text{intra}(\omega)}$. At $\mu=0.01t$ ($n_e
\simeq 7\times10^{10}\,\text{cm}^{-2}$) both the polarizability and
the transverse conductivity peak remain significant. In other words,
one can suppress the amplitude of the Drude peak at lower densities
while not suppressing much the anisotropy and polarizability.

As we pointed out already, by considering the response of an
ensemble of GNRs with fluctuating widths, we are introducing
considerable broadening effects already [compare, for example, the
peak in $\avg{\sigma_{yy}(\omega)}$ for an ensemble in
\Fref{fig:Sigma}, with the five times more intense peak of a single
ribbon (inset)]. For these reasons, we believe that the dichroism of
GNRs remains considerably enhanced in the presence of realistic
moderate disorder. It is worth highlighting also the fact that, since
the dichroism stems here from purely spectral considerations, the
chirality of the GNRs should be immaterial. In fact, all ribbons have
the same scaling of the spectral features with $N$, irrespective of
their chirality, and so we expect the dichroism to remain
when the ensemble comprises GNRs of arbitrary chirality.

Finally, having in mind the scheme depicted in \Fref{fig:Illustration}
where we propose a grating of GNRs, we point out that the dichroism
discussed here so far is intrinsic to each element of the grating, as
it were. This is a departure from the conventional situation where the
grating is made from a normal (isotropic) metal, and the polarizing
effect arises from the geometry only, not from some intrinsic
anisotropy of the metallic comb itself. 
In fact, it might have been noted that, whereas a conventional
metallic grating polarizes \emph{perpendicularly} to the slit
direction, the dichroism of the individual GNRs favors polarization
\emph{along} the ribbon direction. The actual overall polarizing
characteristics of a periodic grating based on GNRs would have to be
determined by the combination of this intrinsic dichroism with the
geometrical effect (just as in a conventional grating), and for which
the surface plasmon-polariton (SPP) physics may play an important role
\cite{Pendry:1999}.
However, SPP excitations contribute to the optical absorption only if:
(i) the incoming wave's frequency coincides with the band where those
excitations are allowed, and not damped;
(ii) the grating is strictly periodic;
(iii) all elements of the grating are metallically connected so as to
maintain coherence of the excitations across the system as a whole;
(iv) the incoming wave impinges the grating at oblique incidence.
Given that we consider only \emph{normal-incidence} (which is the one
typically most straightforward and efficient from an
experimental/applications point of view), the last condition (iv) is
violated from the outset, and corrections to the DP arising from SPP
are not expected. Moreover, one crucial
reason for the existence frequency bands of strong SPP absorption (or
transmission) in 3D metallic gratings arises from the coupling between
those modes at the two opposing surfaces \cite{Pendry:1999}. Being a
strict 2D metallic system (in effect a metallic boundary condition for
the propagation of electromagnetic waves), SPP cannot decay into the
(non-existent) bulk of graphene. This points to the peculiarities of
the SPP physics in this 2D Dirac metal, which have been addressed in
detail in reference~\onlinecite{Bludov:2010}. In particular, this
reference identifies the conditions for the existence of SPP modes,
concluding that they are only allowed in a the range of frequencies
close to the DC limit, where the optical response is dominated by the
Drude peak. Hence, with respect to point (i) above, \emph{even if one
considers the possibility of oblique incidence}, the conditions for
excitation of SPP are rather narrow, and not expected to play a role
at the finite frequencies where the DP effect of
the ribbon system is most effective (see more below). Points (ii) and
(iii) strongly depend on the fabrication process leading to the
ribbons and/or their integration in the final gratings, and are easily
controllable. The main message we wish to underline in this context is
then that, effects associated with increased absorption within certain
frequency bands arising from SPP are not expected in the context of
our proposed setup, and will not influence the DP.
But they could as well be explored by enforcing the conditions
enumerated above, and possibly allow even more versatility and
richness to the polarizing characteristics of nanoribbon-based
gratings. Such considerations are, however, out of the scope of this
report.

%
\section*{Conclusions}

Having derived the exact optical conductivity tensor of GNRs, we
studied the optical absorption response of ensembles of ribbons with
fluctuating width. One verifies that the optical absorption can be
made highly anisotropic within a frequency band that is tunable via
the ribbon average width, and/or via the impedance characteristics of
the embedding medium.
Physically, the origin of such strong anisotropy lies in a resonant
feature that is simultaneously very strong and resilient to level
broadening, in comparison with the conventional van Hove-type
absorption singularities, which quickly wash out in the presence of
width fluctuations and/or disorder.

Quantitative analysis reveals that an ensemble
of monolayer GNRs can show a very high degree of polarization,
$\sim85\%$. This value can be enhanced by placing the ribbon
in a cavity, so that the real part of the impedance is increased in
the appropriate region of the spectrum. In such situations the degree
of polarization can be close to 100\%, which is quite remarkable given
the atomic thickness of the polarizing element.

The current analysis focuses on the intrinsic absorption anisotropy of
GNRs, where disorder effects are mimicked by the fluctuating ribbon
widths. We are currently exploring routes to study the influence of
more specific disorder models, and combining the intrinsic absorption
response of GNRs with the geometric effects expected to arise in a GNR
grating setup. Likewise, the interplay of the anisotropy induced
here by space quantization and plasmons likely to be excited in
such finite-sized geometries should be addressed in the future.

Given the recent developments in precision patterning and growth of
narrow GNRs, and given the technological interest in optical elements
operating in the IR and THz bands, we trust these results can motivate
further theoretical and experimental investigation of GNRs and other
graphene-derived structures towards such applications.

\acknowledgments
We acknowledge insightful discussions with A. H. Castro~Neto and
J. M. B. Lopes~dos~Santos.
FH acknowledges partial support from grant UMINHO/BI/001/2010.
AJC, RMR, MIV, and NMRP acknowledge support from FEDER-COMPETE,
and from FCT grant PEst-C/FIS/UI0607/2011.
VMP acknowledges the support of NRF-CRP award "Novel 2D materials with
tailored properties: beyond graphene" (R-144-000-295-281).

\bibliographystyle{apsrev}
\bibliography{ribbon_dichroism}

\begin{thebibliography}{47}
\bibitem[{End({\natexlab{b}})}]{EndNote-4}
\bibinfo{note}{The situation with nanotubes is quite different, because the
  tubes have no edges, and hence are very weekly affected by disorder. Their
  intrinsic absorption spectrum and the transitions between sub-bands are very
  well reproduced experimentally. But the analogous has not been seen in GNR
  yet, due to the (so far) unavoidable edge roughness.}
\bibitem[{End({\natexlab{d}})}]{EndNote-1}
\bibinfo{note}{For a rectangular cavity, the modes TE$_{10}$ and TE$_{01}$ are
  linearly polarized along the two directions $(a,b)$. For that reason, we need
  a linear superposition of the two to generate an arbitrary linear
  polarization state. The simplest situation is a square waveguide, where the
  two modes are degenerate, and have the same frequency-dependent impedance.}
\end{thebibliography}

\end{document}